# A SURVEY ON DATA WAREHOUSE EVOLUTION

Wided Oueslati[1] and Jalel Akaichi[2]

[1]Department of Computer Science, High Institute of Management, Bouchoucha, Tunisia
`widedoueslati@live.fr`
[2]Department of Computer Science, High Institute of Management, Bouchoucha, Tunisia
`jalel.akaichi@isg.rnu.tn`

## ABSTRACT

*The data warehouse (DW) technology was developed to integrate heterogeneous information sources for analysis purposes. Information sources are more and more autonomous and they often change their content due to perpetual transactions (data changes) and may change their structure due to continual users' requirements evolving (schema changes). Handling properly all type of changes is a must. In fact, the DW which is considered as the core component of the modern decision support systems has to be update according to different type of evolution of information sources to reflect the real world subject to analysis. The goal of this paper is to propose an overview and a comparative study between different works related to the DW evolution problem.*

## 1. INTRODUCTION

Information sources which are integrated in the DW are autonomous and they can change their schema independently of DW. Such changes must be supported when they rich the DW. In fact, the DW technology is seen as the process of good decision making since it provides necessary tools for data analysis such as the On Line Analytical Processing (OLAP). In the literature the DW evolution can be classified into three different approaches namely schema evolution [1, 2, 3, 4] and schema versioning [5, 6] when the DW is defined as a multidimensional schema (fact and dimension tables) and view adaptation and synchronization [7, 8, 9, 10, 11, 12, 13, 14, 15] when the DW is defined as a set of materialized views. The goal of this paper is to present different works related to DW evolution, then to propose a comparative study between those works.

This paper is organized as follows. In section 2, we present different researches works related to DW evolution. In section 3, we present comparative studies between researches works cited above. In section 4, we summarize the work and we propose new perspectives that can be done in the future.

## 2. STATE OF THE ART

In the literature, the DW evolution can be classified into three different approaches namely schema evolution [1, 2, 3, 4], schema versioning [5, 6] and view maintenance [7, 8, 9, 10, 11, 12, 13, 14, 15].

## 2.1. Schema Evolution

This approach focuses on dimensions updates [1, 2], instances updates [4], facts updates and attributes updates [3].

In [1], authors proposed some operators to define dimension updates. Those operators are:

- Generalize operator: it allows the creation of new level (ln) to which a pre-existent one (l) rolls up. Authors took the example of the dimension " store " to which they defined a new level "type of store " then the level " type of store " generalize the dimension "store ".

- Specialize operator: it allows adding a new level (ln) to a dimension. Authors choose to specialize the dimension "day" with the level "hour ", and then the level "hour" specializes the dimension "day ".

- Relate levels operator: it allows defining a roll up function between two independent levels of the same dimension. Authors defined a relation between the level "category" and the level "brand". Those two levels were independent.

- Unrelated levels operator: it allows deleting a relation between two levels. Authors deleted the relation between the levels "company" "category" and the level "brand ".

- Delete level operator: it allows to delete a level then to define new relations between levels. Authors deleted the level "branch" then a direct relation between the levels "category" and "item" was defined.

- Add instance operator: it allows adding an instance to a level in the dimension. Authors added the instance item 5 to the level "item ".

- Delete instance operator: it allows deleting an instance of a level. Authors deleted the instance item 4 of the level "item ".

After defining operators to handle dimension updates, authors of [1] saw that they must handle the impact of dimension structural updates on the data cube. In fact, they proposed some data cube adaptation after the Dellevel update, Addlevel update, DelInstance update, AddInstance update by computing for each cube view an expression to maintain it.

In this paper, no implementation was done to support those changes impact.

In [2] authors proposed an extension to the work presented in [1] and defined the WareHouse Evolution System (WHES) prototype to support dimensions and cubes update. In fact, they extended the SQL language and gave birth to the Multidimensional Data definition Language (MDL). This latter allowed defining operators for to support evolution of dimensions and cubes.

For dimensions update, authors defined the following operators:

- CreateDimension: this operator allows the creation of a new dimension (with its name, its properties and its levels).

- DropDimension: this operator allows the deletion of an existing dimension (with its name, its properties and its levels).

- RenameDimension: this operator allows changing the name of a given dimension.

- AddLevel: this operator allows the add of a new level to a given dimension.

- DeleteLevel: this operator allows the deletion of a level from a given dimension.

- RenameLevel: this operator allows changing the name of a given level.

- AddProperty: this operator allows the add of a property or attribute to a given dimension or a given level.

- DeleteProperty: this operator allows the deletion of a property from a dimension or from a level.

To add a level X (brand) to a dimension Y (product) using the MDL, users should respect the following syntaxes with X= brand and Y= product:

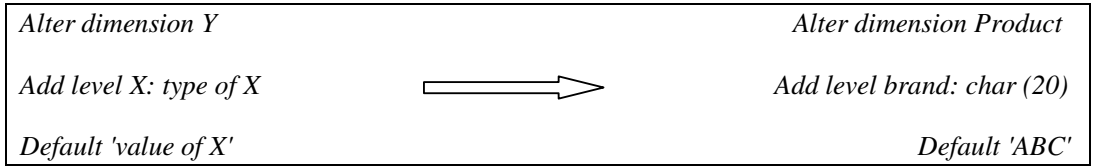

| *Alter dimension Y* | *Alter dimension Product* |
|---|---|
| *Add level X: type of X* | *Add level brand: char (20)* |
| *Default 'value of X'* | *Default 'ABC'* |

For cube updates, authors defined the following operators:

- CreateCube: this operator allows the creation of a new cube.

- DropCube: this operator allows the deletion of a given cube.

- RenameCube: this operator allows the change of the name of a given cube.

- AddMeasure: this operator allows the add of a measure to a given cube.

- DeleteMeasure: this operator allows the deletion of a measure from a given cube.

- RenameMeasure: this operator allows the change of the name of a given measure.

- AddAxis: this operator allows the add of an axis of analyse to a given cube.

- DeleteAxis: this operator allows the deletion of a given axis of analyse from a cube.

Let's mention that a cube is the fact table and the axis is the dimension in the relational schema.

To add an axis X (city) to a cube Z (sales) using the MDL, users should respect the following syntax:

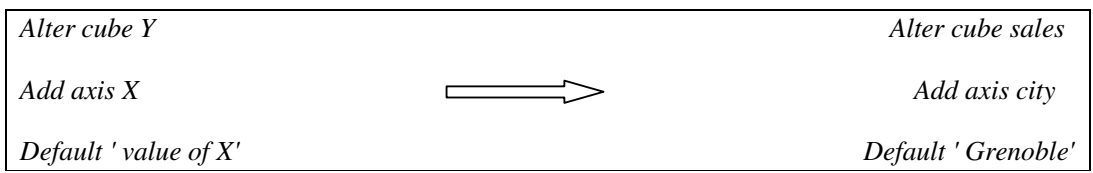

| *Alter cube Y* | *Alter cube sales* |
|---|---|
| *Add axis X* | *Add axis city* |
| *Default ' value of X'* | *Default ' Grenoble'* |

In [3], authors defined a formal description of multidimensional schemas and instances. This formal description constitutes the data model. This latter was defined as follows: a MD model M is a 6 tuple (F, L, A, gran, class, attr) where F is a finite set of fact names, L is a finite set of dimension level names, A is a finite set of attributes names, Gran is a function that associates a fact with a set of dimension level names, Class : is a relation defined on the level name, Attr is a function mapping an attribute to a given fact or to a given dimension level

After defining the data model, authors presented a set of formal evolution operations. Those latter can have an effect on the model or not. The following evolution operations have no effects on the model:

- Insert level: it consists on extending the MD model by a new dimension level. This operation has no effects on instances.

- Delete level: it consists on deleting a dimension level from an MD model but the deleted dimension must not be connected to the fact. This operation has no effects on instances.

- Insert attribute: it consists on creating new attribute without attaching it to a dimension level or fact. This operation has no effects on instances.

- Delete attribute: it consists on deleting an attribute which is a disconnected attribute (A € F, A € D). This operation has no affects on instances.

- Insert classification relationship: it consists on defining a classification relationship between two existing dimension levels. This operation has no effect on instances.

- Delete classification relationship: it consists on deleting a classification relationship without deleting the corresponding dimension levels. This operation has no effect on instance.

The following evolution operations have effects on the model:

- Connect attribute to dimension level: it consists on connecting an existing attribute to a dimension level. This operation has an effect on the instance. In fact, it should define a new function for each new attribute to assign an attribute value to each member of the corresponding level.

- Disconnect attribute from dimension level: it consists on disconnecting an attribute from a dimension level. This operation has an effect on the instance since it should eliminate the deleted attribute functions.

- Connect attribute to fact: it consists on connecting an existing attribute to a fact. This operation has an effect on the instance. In fact, it should define a function that maps coordinates of the cube to measures.

- Disconnect attribute from fact: it consists on disconnecting an existing attribute from a fact. This operation has an effect on instance. In fact, it should delete the function that maps coordinates to measures.

- Insert fact: it consists on extending the MD model by a new fact and without attaching dimension levels to this fact. It should define dimensions for this fact separately. This operation has no effect on the instance but has an effect on the MD model since it should define a new function that associates a fact with a set of dimension level names.

- Delete fact: it consists on removing an existing fact from the MD model but this fact must not be connected to any dimension and don't contain any attributes. This operation has no effect on the instance but has an effect on the MD model since the name of the deleted fact will be removed from the finite set of fact names.

- Insert dimension into fact: it consists on inserting a dimension at a given dimension level into an existing fact. This operation has as an effect the computing of the new fact.

- Delete dimension: it consists on deleting a dimension which is connected to a fact from it. This operation has as an effect the deleting of the function that maps coordinates of the cube to measures.

In [3], authors defined a schema evolution algebra based on formal description but no implementation was done.

## 2.2. Schema Versioning

This approach focuses on keeping trace of changes by keeping different versions of a given DW [5, 6]. In [5] authors make the difference between schema evolution and schema versioning. In fact, for them schema evolution consists in transferring old data from old schema and updating it in a new schema. However, schema versioning consists in keeping the history of all versions by temporal extension or by physical storing of different versions. For the schema versioning, authors of [5] distinguish two types of versions. In fact, they presented real versions and alternative versions. The real versions support changes related to external data sources (changes in the real world) but the alternative versions support changes occurred by the " what if analysis". That happens when decision makers try to predict or to simulate different virtual possible business scenarios. Let's mention that an alternative version is or will be created from a real version and several alternative versions can be created from several real versions or from the same real version. As real version, authors of [5] presented the example of changing the borders of regions (city Konin moved from the region A to the region B) in the case of police DW and its impacts on results of the measure "total-fine" of the fact table "inspected violation". In this case, authors proposed to keep the old version with data before changes and to create a new version with data after changes. As alternative version, authors presented virtual scenarios. They simulated the scenario of moving the violation 2 from the group A to the group B, then the decision maker can compare the real situation with the virtual one. In [5], every version real or alternative has a valid time [begin valid time, end valid time] in which the version is called valid. It is the time constraints on versions.

At the implementation level, authors used the data sharing technique to avoid the physical copy of data in every DW version. In fact, they stored in a given DW version only data that are new or changed in a given version and other data related to a parent version and then shared by its child versions. To model this, a prototype multiversion DW was implemented in visual C++.

In [6], authors handled evolutions in multidimensional structures and provided a new conceptual model. In fact, they proposed a case study in which the fact table is related to birds and dimensions tables are Date, Gender, City and country. Authors defined for each city the districts belonging to it in the year 2001 (the districts D1 and D2 belong to the city C1 and D3 belongs to C2). In 2002, some changes happened. In fact, D2 was not any more belongs to C1 but to C2. Due to this change the query about the number of birds per year and per district will change. To solve this problem, authors of [6] handled dimension schema evolution (creation and deletion of a dimension, creation and deletion of a hierarchy, creation and deletion of a level, move of a level in the hierarchical schema structure) and evolution on members (creation of a member, deletion of a member, transformation of a member's name or attribute, merging of n members into one member, splitting of one member into n members and reclassification of a member in the dimension structure). In fact, for each change, a new version was defined in order to keep trace and to respect the definition of a DW (time variant). Each version is valid within a time valid interval. This solution was developed with the visual basic interface on the commercial OLAP environment.

## 2.3. View Synchronization and Maintenance

This approach focuses on maintaining a materialized view in response to data changes [7, 8, 9, and 10] or to data sources changes [10, 11, 12, 13, and 14] and sometimes to monitor the DW quality under schema evolution [15].

Research works elaborated in the context of view maintenance can be classified in the following categories:

- View adaptation [7, 8, 9, and 10]: this approach consists in adapting views to changes by adding meta data to materialized views. Those meta data contain structural updates related to materialized view.
- View synchronization (rewriting of views) [10, 11, 12, 13, and 14]: many research works were interested in this approach because it is in relation with other problems such as data integration, data warehouse modeling…

In [10] Bel presents an approach for dynamic adaptation of views related to data sources

(relations sources) changes. The main idea of this work is to avoid the recomputation of views which are defined from several sources. In fact, the key idea is to compute the new view from the old one. Bel presents the view adaptation problem from two different points of views.

The first point of view is from the user or from the DW designer or administrator and the second point of view is from data sources. For the first point of view, the user or the DW designer can bring into play schema changes on views (e.g. adding an attribute, delete an attribute, modifying an attribute domain in a view schema) independently of the data sources. Then the changes of the view definition lead to recompute the materialized view. This is the so called "view adaptation". For the second point of view, the data sources (relation sources) can change their schema. This type of changes can touch the DW structural consistency since it may invalidate the materialized views. In this case the solution is to preserve the structural consistency of the DW. This kind of view adaptation is so called "the structural view maintenance". Bel investigated the view adaptation problem from both point of views citing above. Let's start with the view adaptation from point of view data sources or relation sources: the author of [10] presents the impact of schema changes of data sources on the SELECT, WHERE and FROM clauses of the view query.

For the SELECT clause, authors took the case of deleting an attribute A from a data source and its impact on the materialized view V. This latter should be rewritten (V') in order to be updated by deleting the attribute A, then the old view V is deleted and the new view V' is renamed as V. this idea was translated into the following query:

*Create view V' as*

*(Select\* From V)*

*Hide attribute A*

*Drop view V'*

*Rename V' as V*

For the WHERE clause, Bel presents two cases:

- The deleted attribute is involved only in the WHERE clause.
- The deleted attribute is involved in the SELECT and the WHERE clause.

In the first case, the deleted attribute A1 was used in the condition $C_1$ of the view V. the solution is to define a new view V' from the old one V. V' will contain all attributes (*) of V and other attributes from different relation sources ( $R_1$......$R_m$) under different conditions expect the condition $C_1$ which contains the deleted attribute $A_1$. This idea was translated into the following query:

*Create view V' as (Select\*From V)*

*Union (select\* From $R_1$, $R_2$, .....$R_m$*

*Where not $C_1$ and $C_2$ and......and $C_k$);*

In the second case, the deleted attribute A1 is involved in the SELECT and WHERE clause. The solution is to define a new view V'. This latter will contain all attribute of the old view V expect the deleted attribute $A_1$ ( $A_2$,...$A_n$) and other attributes expect $A_1$ from different relation sources ( $R_1$,....$R_m$) under some conditions expect the condition C1 which contains the deleted attribute A1. The old view V will be deleted and replaced by the new view V' since this latter will be renamed as V. This idea was translated into the following query:

*Create view V' as*

*(Select $A_2$, $A_3$,...$A_n$ From V)*

*Union (select $A_2$, $A_3$, ...$A_n$*

*From $R_1$, $R_2$, .....$R_m$*

*Where not $C_1$ and $C_2$ and...$C_k$)*

*Drop view V*

*Rename V' as V*

For the FROM clause, Bel presents two cases of impact of DELETE a relation source on the FROM clause: existence of only one relation source or existence of several relation sources.

In the first case, the delete of the only relation source from the FROM clause entails the delete of the entire view. In the second case, the delete of a relation source $R_1$ will occur the replacement of the old view V by a new one V'. This latter will contain attributes of V and other attributes from different relation sources expect the deleted one of course under some conditions expect the condition which contains an attribute related to the deleted relation source. This idea was translated into the following query:

*Create view V' as*

*(Select $R_2.A_i$, …$R_m.A_j$ From v) Union*

*(Select $R_2.A_i$, …$R_m.A_j$ From $R_2$,…..$R_m$*

*Where not ( $R_1.A_i = R_2.A_i$ ) And ( $R_2.A_i = R_m.A_j$ ))*

As mentioned above, the views can change their schema independently of data sources. In fact, the user or the DW administrator can bring into play schema changes directly on views by adding an attribute to a view or by modifying the domain of an attribute.

In [10], the add of new attribute to a view is the result of new requirements. This operation does not need the rewriting of the view but only the storage of the value of the new attribute. This primitive has the following form:

Add attribute to V (A: <type of A>)

The modification of an attribute domain simulates the creation of new view V' from the old one V. V' must not involve the specific attribute, then an attribute with the same name and having a new type or new definition (domain) is added. The old view V is deleted and the new view V' is renamed as V. The following query shows the modification of the domain (new type T) of the attribute A:

*Create view V' as*

*Select*From V*

*Hide attribute A*

*Add attribute A: T*

*Drop view V*

*Rename V' as V*

EVE system [11] proposes a prototype solution to automate view definitions rewriting to solve the problem of view inflexibility. This solution has the goal to preserve the maximum number of affected view definitions by the occurrence of information sources schema changes. The EVE approach assumes that information sources are integrated in the EVE system via a wrapper which translates their models into a relational common model. They are supposed to be heterogeneous and autonomous which join, or change dynamically their capabilities such as their schema.

EVE system includes two basic modelling tools: a model permitting to user to express view definition evolution via an extended SQL called Evolvable SQL (E-SQL) [11] and a model for the description of the information sources (MISD) [11] and the relationships between them. This model of Information Sources description can be exploited for seeking a suitable substitution for the affected view definition components (attributes, relations, and conditions).

The View Knowledge Base (VKB) described by E-SQL and the Meta Knowledge Base (MKB) revealed by MISD, represent the base for any operation of view rewriting or view synchronization process.

The Data Warehouse Management System constitutes an intermediary between the user space called Data Warehouse and the information space including the participating data sources. When an information source joins the structure the DWMS, it provides its structure, its data model and eventually its content. This information is stored into the MKB with respect to the MISD.

As well, the relationships between information sources, also called substitution rules, can be added by the DWMS administrator and/or generated automatically, then inserted into the MKB. This information constitutes the key platform for finding affected view definitions components substitutions.

Another contribution of EVE approach is to propose an E-SQL language allowing user preferences placing into SQL view definition. E-SQL is an extension of SELECT-FROM-WHERE SQL enriched by specifications defined by the developer in charge of the view definitions in order to indicate how those latter can evolved. The E-SQL defined views are then stored into a structure called View Knowledge Base.

The view synchronization [11] consists in determining legal rewritings for the affected views, referring to the rules or constraints embodied into the MKB. These rules enable substitutions retrieval for the affected view definition components while respecting preference parameters described into the VKB.

The view rewriting is legal when it is compatible with the current information space. This rewriting have to preserve the information presented by the initial view definition according to preferences parameters associated to the view definition components and the possibilities of substitutions offered by the MISD. In [11], the user must intervene in preferences definition otherwise; the system can not take into account the schema changes.

In the context of extension of SQL, the authors on [12] defined the cooperative SQL named C-SQL which detends the simplicity of SQL. In [13], authors defined the schema SQL named S-SQL. This latter allows integrating the relational databases and meta data. This language can be exploited by database analyst in order to describe the schema transformation.

The increase need to decrease the network saturation and to minimize communication costs, have led [14] to the EVE solution to become more adapted for dynamic and distributed environment by adopting new techniques like the mobile agents. In fact, authors of [14] propose to design a mobile agents view synchronization system based on EVE called MAVIE. This latter has to ensure data warehouse maintenance under schema changes.

MAVIE solution decreases the synchronization time due to parallelism permitted by mobile agents and avoids the saturation of the network. The architecture of MAVIE system [14] is distributed on four entities which are the mobile MKB agent, the mobile VKB agent, the mobile detector agent and the mobile synchronizer agent. All those agents know each other via their identifier, names and sites. That fact will assure the direct communication between all mobile agents.

The following figure describes the communication between mobile agents of MAVIE system.

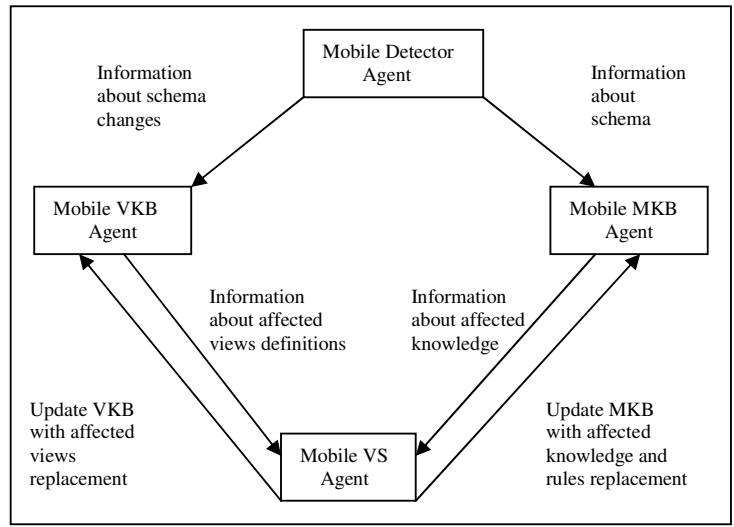

Figure 1: MAVIE agent communication [14]

Mobile information sources can change their schema continuously. This fact makes the views of the data warehouse undefined. To solve this problem, mobile agent technology imposes it self as a new solution to schema change detection system. In fact the mobile detector agent in contrast to the static one can move from one information source to another to detect changes.

The change detection operation consists on comparing for each schema component, two schema versions (schema$_{t-1}$ , schema$_t$ ), if they have been found different, it implies that a change has been occurred and it has to be computed. This latter will be sent in parallel to the mobile Meta Knowledge Base (MKB) and the mobile View Knowledge Base (VKB).

After receiving messages from mobile detectors about schema changes, the mobile MKB agent must look into all MKB components in order to determine affected knowledge and rules by the indicated schema changes, and then it computes the affected knowledge before sending them to the mobile view synchronizer agent.

After receiving messages from mobile detectors about schema changes, the mobile VKB agent must look into all VKB components in order to determine affected views by the indicated schema changes, and then it computes the affected views before sending them to the mobile view synchronizer agent.

The role of the mobile view synchronizer (VS) agent is to find legal rewritings for the affected views (the view rewriting is legal when it is compatible with the current information space), then it emits the affected knowledge and views replacement to the mobile MKB agent and the mobile VKB agent respectively.

The above solution based on EVE and on a technique resulting from the field of artificial intelligence which is the mobile agents called MAVIE permitted more autonomy ( by separating between schema change detection, affected knowledge and view definition determination and view definition synchronization tasks), parallel schema change detection (by using parallel detector agent instances running over distributed information sources), parallel determination of affected knowledge and view definition determination and parallel view definitions synchronization ( by triggering view synchronizer agent instances).

In [15], Quix saw that the quality of the DW is important. In fact, his approach is embedded in the Data Warehouse Quality (DWQ) framework. The aim of [9] is not to provide new techniques to maintain views but to monitor the DWQ under evolution. In fact, Quix presented many evolution operations and their impacts on quality factor. For example, the add /delete of a view and the add/delete of an attribute to/from view will affect the completeness, the correctness and the consistency of the logical schema. The rename of a view will affect the interpretability and understand ability of the view and its attributes. The change of an attribute domain will affect the interpretability of data. The add of an integrity constraint will affect the credibility and consistency of data in data store. The delete of an integrity constraint will affect the consistency of data.

## 3. COMPARATIVE STUDIES

In this section we propose to compare researches works related to the same approach of DW evolution. For the schema evolution approach, our comparative study is based on the following criteria: level evolution, instance evolution, dimension evolution and fact evolution.

Works in [1], [2] and [3] supported the level evolution but if we detail this type of evolution we will find that each author supported a set of level evolution but not the total. In fact, authors in [1], [2] and [3] tacked into account the Add and the Delete of level but only authors in [1] treated the Relate and Unrelate of level and only [2] treated the Rename of level.

Hurtado in [1] was interested in Instance evolution by introducing the Add and Delete of instance.

Benitez [2] and Blashka [3] were interested in Dimension evolution. In fact, both of them presented the Add, the Delete of dimension and the Add, the Delete of attribute but only [2] presented the rename of dimension.

Benitez [2] and Blashka [3] were interested in Fact evolution but each author was interested in some operators which supported such evolution. In fact, Benitez [2] presented the Add, the Delete and the Rename of measure while Blashka presented the Add and the Delete of fact.

To summarize this section we propose the following table (table 1):

Table1. Schema evolution

| Criteria<br><br>Author | Level<br><br>evolution | Instance<br><br>evolution | Dimension<br><br>evolution | Fact<br><br>evolution |
|---|---|---|---|---|
| Hurtado [1] | x | x | | |
| Benitez [2] | x | | x | x |
| Blaschka [3] | x | | x | x |

Dealing with the schema versioning, we can base our comparative study on the following criteria: real version and alternative version.

As we mentioned above, the real version supports changes of the real world while the alternative version supports the what-if-analysis since it is used to simulate different business scenarios. Authors in [5] and [6] were interested in presenting real versions. Authors in [6] simulated some business scenarios by presenting alternative versions.

To summarize this section we propose the following comparative table (table 2):

Table2. Schema versioning

| Criteria<br><br>Author | Real versions | Alternative Versions<br><br>(what if analysis) |
|---|---|---|
| Bebel [5] | x | |
| Body [6] | x | x |

For the view maintenance, we can base our comparative study on the following criteria: view adaptation, view synchronization, extension of SQL and impact of schema changes on DWQ.

The user or the DW administrator can bring modification directly on views independently of data sources by adding attributes or relations to a view definition. This is the so called view adaptation. The data sources can change their schema and this may invalidate the materialized views. This is the so called the structural view maintenance.

Authors in [10], [11], [12], [13] and [14] focuses on structural view maintenance. In fact, they presented the problem of DW evolution when the data sources change their schema.

Authors in [7], [8], [9] and [10] treated the DW evolution problem when data change independently of data sources changes so they dealt with the view adaptation approach. Jalel in [14] was the only author that used a new technique based on mobile agents to decrease the synchronization time and then to avoid the saturation of the network.

Authors in [11], [12] and [13], defined new languages based on SQL. In fact, authors in [11] defined the E-SQL, authors in [12] defined the C-SQL and authors in [13] defined the S-SQL.

All those authors cited above focused on view maintenance after data changes or data sources changes but they didn't take into account the quality of the DW, that means the impact of such changes on the data warehouse quality (DWQ) while this was treated by Quix in [15].

To summarize this section we propose the following comparative table (table 3):

Table3. View maintenance

| Criteria<br>Author | View adaptation | View synchronization | Extension of SQL | DWQ |
|---|---|---|---|---|
| Gupta [7] | x | | | |
| Nica [8] | x | | | |
| Rundensteiner [9] | x | | | |
| Bel [10] | x | x | | |
| Zhang [11] | | x | x ( E-SQL) | |
| Rajarman [12] | | x | x (C-SQL) | |
| Lakshmanan [13] | | x | x (S-SQL) | |
| Jalel [14] | | x | | |
| Quix [15] | | | | x |

## 3. CONCLUSION

In this paper, we presented different research works which focused on the DW evolution. Those works were classified into three approaches. In fact, we presented the schema evolution approach, the versioning approach and the view maintenance approach. As future work, we propose to investigate the problem of schema changes in the case of spatio-temporal and trajectory data warehouse.

### ACKNOWLEDGEMENTS

We would like to thank the High Institute of Management of Tunis especially my supervisor Dr Akaichi Jalel and my familly especially my father Oueslati Abdellatif, my mother Ben Moussa Aicha, my sisters Wafa and Amira and my brothers Wissem and Slim for their support !